\documentclass[draft]{agujournal2019}
\usepackage{url} 
\usepackage[inline]{trackchanges} 
\usepackage{soul}
\usepackage{latexsym }
\usepackage{amsmath}
\AtBeginDocument{}

\draftfalse

\journalname{Space Weather}

\begin{document}

\title{Prediction of the SYM-H Index 
Using a Bayesian Deep Learning Method
with Uncertainty Quantification}

\authors{Yasser Abduallah\affil{1,2}, 
Khalid A. Alobaid\affil{1,2,3},
Jason T. L. Wang\affil{1,2} \\
Haimin Wang\affil{1,4,5},
Vania K. Jordanova\affil{6}, 
Vasyl Yurchyshyn\affil{5} \\
Huseyin Cavus\affil{7,8},
Ju Jing\affil{1,4,5}
}

\affiliation{1}{Institute for Space Weather Sciences, New Jersey Institute of Technology, 
Newark, NJ 07102, USA}

\affiliation{2}{Department of Computer Science, New Jersey Institute of Technology, 
Newark, NJ 07102, USA}

\affiliation{3}{College of Applied Computer Sciences, King Saud University, Riyadh 11451, Saudi Arabia}

\affiliation{4}{Center for Solar-Terrestrial Research, New Jersey Institute of Technology,
Newark, NJ 07102, USA}

\affiliation{5}{Big Bear Solar Observatory, New Jersey Institute of Technology, 
Big Bear City, CA 92314, USA}

\affiliation{6}{Space Science and Applications, Los Alamos National Laboratory, Los Alamos, NM 87545, USA}

\affiliation{7}{Department of Physics, Canakkale Onsekiz Mart University, 17110 Canakkale, Turkey}

\affiliation{8}{Harvard-Smithsonian Center for Astrophysics, 60 Garden Street, Cambridge, MA 02138, USA}

\correspondingauthor{Jason T. L. Wang}{wangj@njit.edu}

\begin{keypoints}
\item 
SYMHnet is a novel deep learning method for making short-term predictions of the SYM-H 
index (1 or 2 hours in advance).
\item With Bayesian inference, SYMHnet can
quantify both aleatoric (data) and epistemic (model) uncertainties in making its prediction. 
\item SYMHnet generally performs better than
related machine learning methods for SYM-H forecasting.
\end{keypoints}

\begin{abstract}
We propose a novel deep learning framework, named SYMHnet, which employs a graph neural network and 
a bidirectional long short-term memory network to cooperatively learn patterns from 
solar wind and interplanetary magnetic field parameters
for short-term
forecasts of the SYM-H index based on
1-minute and 5-minute resolution data. 
SYMHnet takes, as input, the time series of the parameters' values
provided by NASA's Space Science Data Coordinated Archive
and predicts, as output, 
the SYM-H index value
at time point $t$ + $w$ hours
for a given time point $t$ 
where $w$ is 1 or 2.
By incorporating Bayesian inference into the learning framework, 
SYMHnet can quantify both aleatoric (data) uncertainty and
epistemic (model) uncertainty when predicting future SYM-H indices.
Experimental results show that
SYMHnet works well at quiet time and storm time,
for both 1-minute and 5-minute resolution data.
The results also show that
SYMHnet generally performs better than related machine learning methods.
For example, SYMHnet achieves a forecast skill score (FSS) of
0.343
compared to the FSS of 0.074 of a recent gradient boosting machine (GBM) method
when predicting SYM-H indices (1 hour in advance) 
in a large storm (SYM-H = $-393$ nT) using 5-minute resolution data.
When predicting the SYM-H indices (2 hours in advance)
in the large storm,
SYMHnet achieves an FSS of
0.553 compared to the FSS of
0.087
of the GBM method.
In addition, SYMHnet can provide 
results for both data and model uncertainty quantification, 
whereas the related methods cannot.
\end{abstract}

\section*{Plain Language Summary}
In the past several years, machine learning and its subfield, deep learning, have attracted considerable interest. Computer vision, natural language processing, and social network analysis make extensive use of machine learning algorithms. Recent applications of these algorithms include the prediction of solar flares and the forecasting of geomagnetic indices. In this paper, we propose an innovative machine learning method that utilizes a graph neural network and a bidirectional long short-term memory network to cooperatively learn patterns from 
solar wind and interplanetary magnetic field parameters to provide 
short-term
predictions of the SYM-H index. In addition, we present techniques for quantifying both data and model uncertainties in the output of the proposed method.

\section{Introduction}\label{sec:introduction}
Geomagnetic activities and events are known
to have a substantial impact on the Earth. 
They can damage and affect technological systems such as telecommunication networks, 
power transmission systems, and spacecraft
\cite{AyalaEffectOfMagneticOnEarth,Vania2020BookRing}.
These activities are massive and scale on orders of magnitude~\cite{NewellOrderOfMag}.
It may take a few days to recover from the damage, depending on its severity.  
These activities and events cannot be ignored regardless of whether they are in regions at high, medium, or low latitudes
\cite{CarterEtAl2016,GaunteAl2007,Modldiwn2016,Tozzi2019,
Viljanen2014}. 
Therefore, several activity indices have been developed to measure the intensity of the geomagnetic effects.
These indices characterize the magnitude of the disturbance over time.
Modeling and forecasting these geomagnetic indices have become a crucial area of study in space weather research. 

Some indices, such as Kp, describe the overall level of geomagnetic activity while others,
such as the disturbance storm time (Dst) index \cite{Vania2016SpWea..14..668W},
describe a specific area of geomagnetic activity.
The Dst index 
has been used to classify a storm based on its intensity
~\cite{Bala2012KpForecasting,GCS2018,LazzusDstSwarmOptizmizedRNN2017,SVMwithDistanceCorrLU201648,DstBayesianXU20233882}.
The storm is intense when Dst is less than $-100$ nT, moderate when Dst is between $-100$ nT and $-50$ nT, 
and weak when Dst is greater than $-50$ nT \cite{GCS2018,DSTNuraeni_2022}.
Another important index is the symmetric H-component index (SYM-H),
which is used to represent the longitudinally
symmetric disturbance of the intensity of the ring current
during geomagnetic storms. 
The SYM-H index is the one-minute version of the DST index, obtained by data from more stations
\cite{Rangarajan1989Geoma...3..323R,
SYMHForecasting2021ComparisonLSTMCNN,
SpatialGStorm2019JGRA..124..982V,
HighResSYMH2006JGRA..111.2202W}.
On the other hand, ASY-H (the asymmetric geomagnetic disturbance of the horizontal component) is quantified as the longitudinally asymmetric part of
the geomagnetic disturbance field at low latitude to midlatitude. 
In addition, there are other indices that can be used to measure the activity of the storm as described in \citeA{Mayaud1980}.

A lot of efforts have been devoted to developing strategies to alleviate the geomagnetic effects on technologies and humans, but it is almost impossible to offer complete protection from the effects \cite{SYMHForecasting2021ComparisonLSTMCNN}. 
Some of these strategies are to predict
the occurrence and intensity of geomagnetic storms
to offer some level of mitigation of their damaging effects. 
For example,
\citeA{BurtonEaAl1975} established an empirical connection between interplanetary circumstances and Dst using a linear forecasting model.
\citeA{Temerin2002Dst} 
developed an explicit model to predict Dst on the basis of solar wind data for the years 1995–1999, by finding functions and values of free parameters that minimize the
root square error (RMS error) between their model and the measured Dst.
\citeA{Wang2033DiffEqu} used differential equation models to examine the effect of the dynamic pressure of the solar wind on the decay and injection of the ring current.
\citeA{YurchyshynBz2004SpWea22001Y} proposed that the hourly averaged magnitude of the \textit{Bz} component of the magnetic field in interplanetary ejecta is correlated with the speed of the CME, 
which may open a way to predict the Dst index using CME parameters.
\citeA{AyalaEffectOfMagneticOnEarth} performed predictions of global magnetic disturbance in near-Earth space in a case study for the Kp index using Nonlinear AutoRegressive with eXogenous (NARX) models. 
Due to the intrinsic complex response of the circumterrestrial environment to changes in the interplanetary medium, 
these simple models were unable to properly and 
fully depict the evolution of the solar wind-magnetosphere-ionosphere system~\cite{Consolini2001,Klimas1996NonLinear,SYMHForecasting2021ComparisonLSTMCNN}. 
To surpass the limitations of simple models and acquire the complex response of the magnetosphere, researchers
resorted to more advanced models such as artificial neural networks (ANNs).

The use of ANNs focused on the prediction of the Dst and Kp indices.
\citeA{Lundstedt-1994} 
constructed the first Dst prediction model employing a time-delay ANN with solar wind parameters as input variables. 
\citeA{LazzusDstSwarmOptizmizedRNN2017}~created a particle swarm optimization method to train
ANN connection weights to improve the accuracy of the prediction of the Dst index. \citeA{Bala2012KpForecasting} combined ANNs and physical models with solar wind and interplanetary magnetic field parameters such as
velocity, interplanetary magnetic field (IMF) magnitude, and clock angle. 
\citeA{ChandorkarGaussianDST2017} used Gaussian processes (GP) to build an autoregressive model to predict the Dst index 1 hour in advance based on the past solar wind velocity, the IMF component $B_{z}$, and the values of the Dst index. This method generated a predictive distribution rather than a single prediction point. However, the mean values of the estimations are not as accurate as those generated by ANNs.
\citeA{GCS2018} overcame the poor performance of GP and constructed a Dst index estimation model by merging GP with a long short-term memory (LSTM) network to obtain more accurate results.
More recently, 
\citeA{DstBayesianXU20233882} 
developed a new GP regression model that performed better than related 
distance correlation learning methods \cite{SVMwithDistanceCorrLU201648}
in forecasting
the Dst index
during intense geomagnetic storms.
\citeA{Raster2013ComparisonDst1MinuteSpWea..11..187R} compared the effectiveness of 
30 Dst forecast models and found that none of the models performed consistently the best for all events. 

Relatively few researchers have focused on the prediction of SYM-H.
This happens probably because of
the high temporal resolution of 1 minute for the SYM-H index, 
which gives rise to a more difficult problem in estimating SYM-H due to its highly oscillating nature \cite{SYMHForecasting2021ComparisonLSTMCNN}.
However, some SYM-H index prediction techniques have been reported in the literature.
\citeA{Caiangeo-28-381-2010} presented the first 5-minute average estimates
of the SYM-H index
throughout large storms between 1998 and 2006 using a NARX neural network with IMF and solar wind data.
\citeA{SYMHForecastStPatric2019JSWSC...9A..12B}
predicted both the SYM-H and ASY-H indices for solar cycle 24
by employing the NARX neural network in a similar way. 
Both \citeA{SYMHForecastStPatric2019JSWSC...9A..12B} and \citeA{Caiangeo-28-381-2010} used the IMF magnitude ($B$), 
$B_{y}$ and $B_{z}$ components, 
as well as the density and velocity of the solar wind as input data for their models.
\citeA{SYMHForecasting2021ComparisonLSTMCNN} provided a comprehensive examination of two well-known deep learning models, namely long short-term memory (LSTM) and a convolutional neural network (CNN), with an average temporal resolution of 5 minutes for the estimation of SYM-H index values (1 hour in advance). 
The authors used the IMF component $B_{z}$,
squared values of the magnitude of the IMF $B$ and the $B_{y}$ component,
measured at L1 by the ACE satellite in GSM coordinates. 
\citeA{2021ColladoSMYH_ASYH_CNN_LSTMForecasting} created neural network models for the SYM-H and ASY-H predictions by combining CNN and LSTM. The authors considered 42 geomagnetic storms between 1998 and 2018
for model training, validation, and testing purposes.
\citeA{2022XGBoostSYMHbyIong} developed a model using gradient boosting machines to predict the SYM-H index (1 and 2 hours in advance) with a temporal resolution of 5 minutes.

In this paper, we present a new method, named SYMHnet, that utilizes cooperative learning of a graph neural network (GNN) and a bidirectional long short-term memory (BiLSTM) network with Bayesian inference to conduct short-term 
(1 or 2 hours in advance)
predictions of the SYM-H index for solar cycles 23 and 24.
We consider temporal resolutions of 1 minute and 5 minutes, respectively, for the SYM-H index.
To our knowledge, this is the first time that 1-minute resolution data have been used to predict the SYM-H index.
Furthermore, our method can quantify both model and data uncertainties when producing prediction results,
whereas related machine learning methods cannot.

The remainder of this paper
is organized as follows.
Section \ref{sec:data} describes the data, including the solar wind
and IMF parameters, as well as geomagnetic storms, used in this study.
Section~\ref{sec:methodology} presents the
methodology, explaining the
SYMHnet framework, its architecture, and the uncertainty quantification algorithm. 
Section~\ref{sec:results} 
evaluates the performance of SYMHnet on 1-minute and 5-minute resolution data.
We also report the experimental results obtained by comparing SYMHnet
with related machine learning methods on 5-minute resolution data.
Section \ref{sec:conclusion} presents a discussion and concludes the paper.

\section{Database}\label{sec:data}
In training and evaluating SYMHnet, we built a database
that combines the solar wind and IMF parameters with the geomagnetic storms studied in this paper. 
This database contains 42 storms selected from the past two solar cycles
(\#23 and \#24).
The storms occurred between
1998 and 2018. 

\subsection{Solar Wind and IMF Parameters}\label{sec:solarwindandimfparameters}
We consider seven solar wind, IMF, 
and derived
parameters:
IMF magnitude ($B$), 
$B_y$ and $B_z$ components, flow speed, proton density, electric field
and flow pressure.
These parameters have been used in related studies
\cite{SYMHForecastStPatric2019JSWSC...9A..12B,Caiangeo-28-381-2010,VaniImprovedEmpiricalSolarWindws2016SpWea..14..511D,2022XGBoostSYMHbyIong}.
 The parameters' values along with the SYM-H index values are collected from the NASA Space Science Data Coordinated Archive available at \url{https://nssdc.gsfc.nasa.gov}~\cite{OMNIWebData}. 
 Data are collected with 1- and 5-minute resolutions. 

\subsection{Geomagnetic Storms}\label{sec:geomagneticstorms}
We work with the same storms as those considered in previous studies
\cite{2021ColladoSMYH_ASYH_CNN_LSTMForecasting,
2022XGBoostSYMHbyIong,
SYMHForecasting2021ComparisonLSTMCNN}.
Table \ref{tab:trainingstorms}
lists the storms used to train SYMHnet.
Table \ref{tab:validationstorms}
lists the storms used to validate SYMHnet.
Table \ref{tab:teststorms} 
lists the storms used to test SYMHnet.
The training set, validation set, and test set are disjoint.
Thus, SYMHnet can make predictions on storms that
it has never seen during training.
Note that each storm period listed in Tables \ref{tab:trainingstorms},
\ref{tab:validationstorms}, and
\ref{tab:teststorms}
contains both quiet time and storm time,
as indicated by the maximum SYM-H and minimum SYM-H values
in the period.

\begin{table}
 \caption{Storms Used to Train SYMHnet}  
 \label{tab:trainingstorms}
 \centering
 \begin{tabular}{c c c c c c}
 \hline
  Storm \# & Start Date & End Date & Min SYM-H (nT) & Max SYM-H (nT) \\
 \hline
1 & 02/14/1998 & 02/22/1998 & $-119$ & 12 \\
2 & 08/02/1998 & 08/08/1998 & $-168$ & 25 \\
3 & 09/19/1998 & 09/29/1998 & $-213$ & 8 \\
4 & 02/16/1999 & 02/24/1999 & $-127$ & 28 \\
5 & 10/15/1999 & 10/25/1999 & $-218$ & 42 \\
6 & 07/09/2000 & 07/19/2000 & $-335$ & 76 \\
7 & 08/06/2000 & 08/16/2000 & $-235$ & 10 \\
8 & 09/15/2000 & 09/25/2000 & $-196$ & 43 \\
9 & 11/01/2000 & 11/15/2000 & $-174$ & 43 \\
10 & 03/14/2001 & 03/24/2001 & $-165$ & 22 \\
11 & 04/06/2001 & 04/16/2001 & $-275$ & 32 \\
12 & 10/17/2001 & 10/22/2001 & $-210$ & 37 \\
13 & 10/31/2001 & 11/10/2001 & $-313$ & 43 \\
14 & 05/17/2002 & 05/27/2002 & $-113$ & 101 \\
15 & 11/15/2003 & 11/25/2003 & $-488$ & 10 \\
16 & 07/20/2004 & 07/30/2004 & $-208$ & 32 \\
17 & 05/10/2005 & 05/20/2005 & $-302$ & 64 \\
18 & 04/09/2006 & 04/19/2006 & $-110$ & 24 \\
19 & 10/09/2006 & 12/19/2006 & $-206$ & 39 \\
20 & 03/01/2012 & 03/11/2012 & $-149$ & 49 \\
 \hline
 \multicolumn{2}{l}{}
 \end{tabular}
\end{table}

\begin{table}
\caption{Storms Used to Validate SYMHnet}
 \label{tab:validationstorms}
 \centering
 \begin{tabular}{c c c c c c}
 \hline
  Storm \# & Start Date & End Date & Min SYM-H (nT) & Max SYM-H (nT) \\ 
 \hline
21 & 04/28/1998 & 05/08/1998 & $-268$ & 50 \\
22 & 09/19/1999 & 09/26/1999 & $-160$ & 64 \\
23 & 10/25/2003 & 11/03/2003 & $-427$ & 33 \\
24 & 06/18/2015 & 06/28/2015 & $-207$ & 77 \\
25 & 09/01/2017 & 09/11/2017 & $-144$ & 54 \\
 \hline
 \multicolumn{2}{l}{}
 \end{tabular}
\end{table}

\begin{table}
 \caption{Storms Used to Test SYMHnet}
 \label{tab:teststorms}
 \centering
 \begin{tabular}{c c c c c c}
 \hline
  Storm \# & Start Date & End Date &  Min SYM-H (nT) & Max SYM-H (nT) \\ 
 \hline
26 & 06/22/1998 & 06/30/1998 & $-120$ & 39 \\
27 & 11/02/1998 & 11/12/1998 & $-179$ & 19 \\
28 & 01/09/1999 & 01/18/1999 & $-111$ & 9 \\
29 & 04/13/1999 & 04/19/1999 & $-122$ & 63 \\
30 & 01/16/2000 & 01/26/2000 & $-101$ & 21 \\
31 & 04/02/2000 & 04/12/2000 & $-315$ & 16 \\
32 & 05/19/2000 & 05/28/2000 & $-159$ & 47 \\
33 & 03/26/2001 & 04/04/2001 & $-434$ & 109 \\
34 & 05/26/2003 & 06/06/2003 & $-162$ & 10 \\
35 & 07/08/2003 & 07/18/2003 & $-125$ & 23 \\
36 & 01/18/2004 & 01/27/2004 & $-137$ & 41 \\
37 & 11/04/2004 & 11/14/2004 & $-393$ & 92 \\
38 & 09/10/2012 & 10/05/2012 & $-138$ & 18 \\
39 & 05/28/2013 & 06/04/2013 & $-134$ & 37 \\
40 & 06/26/2013 & 07/04/2013 & $-110$ & 19 \\
41 & 03/11/2015 & 03/21/2015 & $-233$ & 62 \\
42 & 08/22/2018 & 09/03/2018 & $-205$ & 26 \\
 \hline
 \multicolumn{2}{l}{}
 \end{tabular}
\end{table}

\section{Methodology}\label{sec:methodology}

Machine learning (ML) and its subfield, 
deep learning (DL) \cite{Goodfellow_DeepLearningBookDBLP:books/daglib/0040158}, have been used extensively in the space weather community for predicting solar flares \cite{MLaaS_RAA_Abduallah_2021,DeepLearningFlareForecastingLoS2018ApJ...856....7H,Liu_2019FlarePrediction}, flare precursors 
\cite{CMH-2019}, 
coronal mass ejections \cite{KhalidCMEFronteirs2022,Liu_2020CMEPrediction}, 
solar energetic particles 
\cite{AbduallahSEP2022, 2009SpWea...7.4008L, 2021SoPh..296..107L,
SEPE10MeV2011,
2021SpWea..1902794S}, 
and geomagnetic indices
\cite{2008JASTP..70..496A,
Bala2012KpForecasting,SYMHForecastStPatric2019JSWSC...9A..12B,2021ColladoSMYH_ASYH_CNN_LSTMForecasting,GCS2018,LazzusDstSwarmOptizmizedRNN2017,2006SPIMFAnGeo..24..989P,SYMHForecasting2021ComparisonLSTMCNN}.
Different from the existing methods,
SYMHnet combines a graph neural network (GNN) and
a bidirectional long short-term memory 
(BiLSTM) network to jointly learn 
patterns from input data.
GNN learns the relationships among the parameter values in the input data, while BiLSTM captures the temporal dynamics of the input data.
As our experimental results show later, this combined learning framework
works well and generally performs better than 
related machine learning methods 
for SYM-H index forecasting.

\subsection{Parameter Graph}
\label{sec:graphrepresentation}
\begin{figure}
    \centering
    \includegraphics[width=1\textwidth]{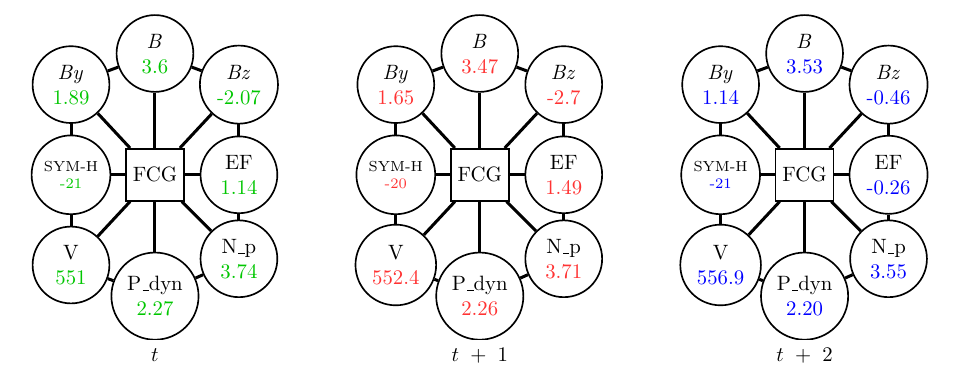}
    \caption{Illustration of
    the parameter graphs constructed at time points
  $t$, $t$ + 1, $t$ + 2, respectively with a
 resolution of 1 minute
 for predicting the SYM-H index 1 hour in advance.
 Each graph contains seven parameters:
 IMF magnitude ($B$), 
$B_{y}$ component,
$B_{z}$ component,
electric field (EF),
proton density (N\_p),
flow pressure (P\_dyn), and
flow speed (V).
    The colored values in the graphs represent the parameters' values that change as time goes on, while the topologies of the graphs remain the same. 
    The value in the SYM-H node in a graph is the label of the graph.
    The FCG symbol in a graph indicates that the graph is fully connected.}
    \label{fig:gnnstaticstructure}
    \vspace{-0.8cm}
\end{figure}

We construct an undirected unweighted fully connected graph (FCG) 
for the solar wind, the IMF and the derived parameters considered in this study,
where each node represents a parameter and there is an edge between every two nodes.
Because the parameter values are time series,
we obtain a time series of parameter graphs
where the topologies of the graphs are the same, but
the node values vary as time goes on.
For example, Figure~\ref{fig:gnnstaticstructure} shows three parameter graphs constructed at time points $t$, $t$ + 1, $t$ + 2, respectively, with a resolution of 1 minute to predict the SYM-H index 1 hour in advance.
 In Figure~\ref{fig:gnnstaticstructure},
 the leftmost graph at $t$
 contains the values of the seven parameters,
 represented by seven nodes or circles, at the time point $t$.
 The FCG symbol in the center indicates that this is a fully connected graph
 in which every two nodes are connected by an edge.
 (For simplicity, only a portion of the edges are shown in the figure.)
 Furthermore, the graph contains a node that represents the value of the SYM-H index at the time point $t$ + 1 hour.
 During training,
this SYM-H index value is used as the label for the graph.
The GNN in SYMHnet will learn the relationships
among the parameters' values and the relationships
between the parameters' values and the label.
If we want to predict the SYM-H index
2 hours in advance, 
then the label will be the SYM-H index value at the time point
$t$ + 2 hours.

The middle graph at $t$ + 1
in Figure~\ref{fig:gnnstaticstructure}
contains the values of the seven parameters at the
time point $t$ + 1 minute.
In addition, this graph contains the SYM-H index
value at the time point
($t$ + 1 minute) + 1 hour,
which is the label for this graph.
If we want to predict the SYM-H index
2 hours in advance, 
then the label will be the SYM-H index value at the
time point
($t$ + 1 minute) + 2 hours.

The rightmost graph at
$t$ + 2 
in Figure~\ref{fig:gnnstaticstructure}
contains the values of the seven parameters at the
time point
$t$ + 2 minutes.
Additionally, this graph contains the SYM-H index
value at the time point
($t$ + 2 minutes) + 1 hour,
which is the label for this graph.
If we want to predict the SYM-H index
2 hours in advance, 
then the label will be
the SYM-H index value at the
time point
($t$ + 2 minutes) + 2 hours.

During testing/prediction, given the values of the seven parameters at
a time point $t'$ (without a label), 
SYMHnet will predict the label, which is the SYM-H index value at the time point 
$t'$ + 1 hour (for 1-hour ahead predictions) 
or the SYM-H index value at the time point $t'$ + 2 hours 
(for 2-hour ahead predictions), 
as detailed in Section \ref{sec:architecture}.

\begin{figure}
    \centering
        \includegraphics[width=0.95\textwidth]{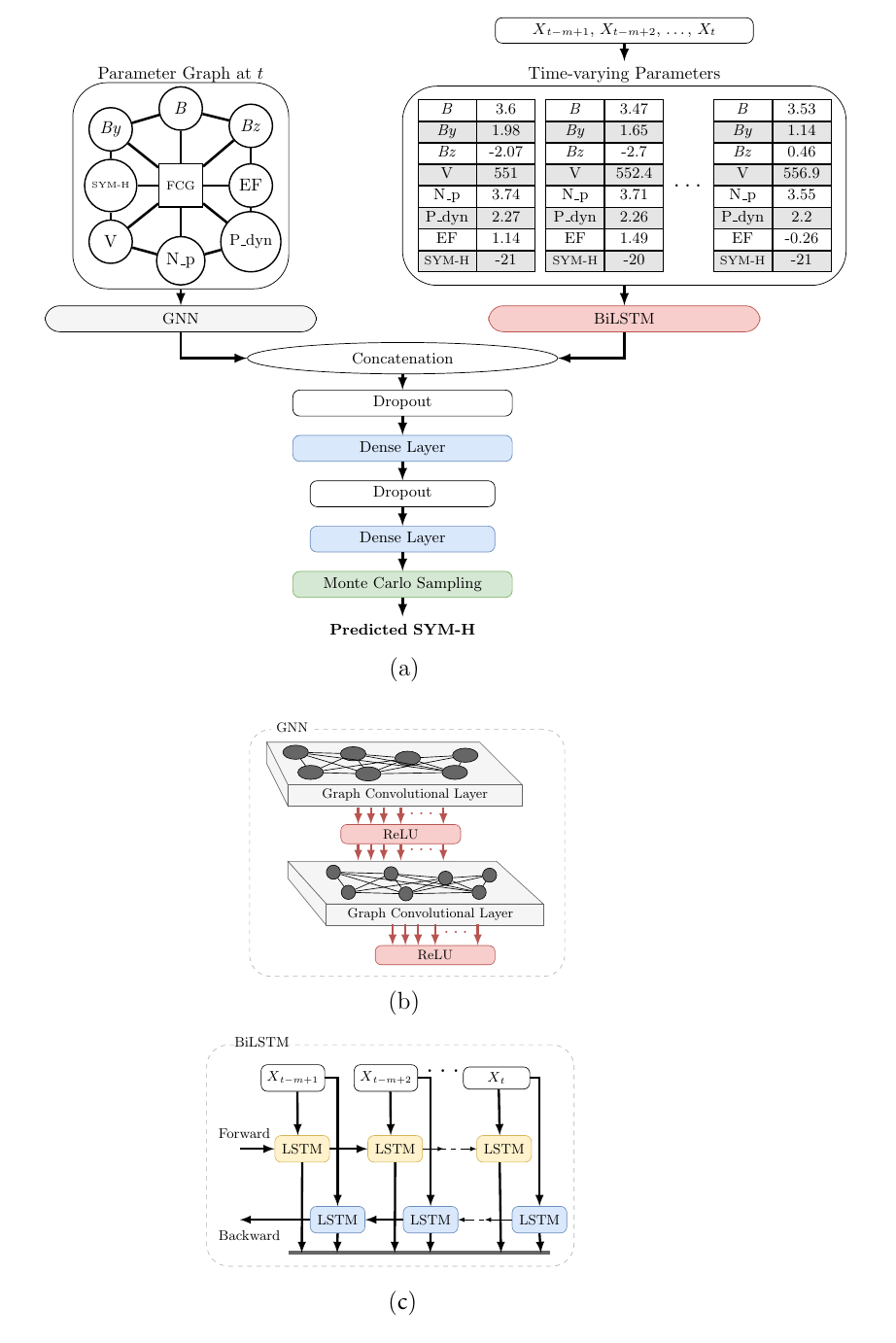}
    \caption{The SYMHnet framework:
    (a) the overall architecture of SYMHnet, 
    (b) the architecture of its GNN component, 
    and (c) the architecture of its BiLSTM component.
    The input parameter graph is for illustration; the actual graph in the implementation is a fully connected graph (FCG). 
    \textit{B} = IMF magnitude (B), \textit{By} = \textit{By} component, \textit{Bz} = \textit{Bz} component, 
    EF = Electric field, 
    N\_p = Proton density, P\_dyn = Flow pressure, and V = Flow speed.}
    \label{fig:architecture}
\end{figure}

\subsection{The SYMHnet Framework}
\label{sec:architecture}
Figure~\ref{fig:architecture} illustrates the SYMHnet framework. 
During training, 
we feed the input data sample at each time point in turn to SYMHnet.
The input data sample at the time point $t$
consists of the parameter graph $G_{t}$
constructed at $t$
and a sequence of $m$ records
$X_{t-m+1}$, $X_{t-m+2}$, $\ldots$, $X_{t}$
where $X_{i}$, $t-m+1 \leq i \leq t$, represents the record collected at the time point $i$.
$X_{i}$ contains the seven values of the solar wind and IMF parameters
along with the SYM-H index value
at the time point $i$.
Including previous SYM-H index values in the input
to predict future SYM-H indices improves prediction accuracy \cite{2022XGBoostSYMHbyIong}.
The number of records, $m$, in the input is set to 10 which was determined by our experiments. When m $<$ 10, BiLSTM cannot effectively capture the temporal patterns in the data. When m $>$ 10, it causes additional overhead for larger sequence sizes without improving prediction accuracy.
The label of the graph $G_{t}$
is used as the label of the input data sample
at the time point $t$.

The parameter graph $G_{t}$ is sent to SYMHnet's GNN component \cite{TGNNCovid2021}
while the sequence of $m$ records, $X_{t-m+1}$, $X_{t-m+2}$, $\ldots$, $X_{t}$, is sent to
SYMHnet's BiLSTM component \cite{AbduallahSEP2022}.
The GNN, illustrated in Figure~\ref{fig:architecture}(b),
contains a graph convolutional layer
followed by a rectified linear unit (ReLU),
which is followed by another graph convolutional layer and ReLU.
The BiLSTM network,
illustrated in Figure~\ref{fig:architecture}(c),
is composed of two LSTM layers \cite{LSTMHochreiter1997LongSM} with opposite directions when processing the data. 
This architecture allows the BiLSTM network to use one LSTM layer to read the sequence from the end to the beginning, denoted as forward, and the other LSTM layer to read the sequence from the beginning to the end, denoted as backward.
GNN is good for learning the correlations between
nodes (parameters) in a graph \cite{TGNNCovid2021}
while BiLSTM is suitable for learning the temporal patterns
in time series \cite{AbduallahSEP2022,LSTMbiLSTMComparison2019}.
SYMHnet combines the learned parameter correlations and temporal patterns
into a joint pattern, which is then passed
to two dropout and dense layers.

A dropout layer provides a mechanism to randomly drop a percentage of neurons to avoid over-fitting 
on the training data so that the SYMHnet model can generalize
to unseen test data.
It also enables the Monte Carlo (MC) sampling method
described in Section \ref{sec:uq}
because the internal structure of the network is
slightly different each time neurons are dropped \cite{MoteCarolDropout10.5555/3045390.3045502, HaodiFibri2021}.
Each neuron in a dense layer connects to every neuron in the preceding layer
\cite{Goodfellow_DeepLearningBookDBLP:books/daglib/0040158}.
The dense layer helps to change the dimensionality of the output of the preceding layer
so that the SYMHnet model can easily define the relationship
between the values of the data on which the model works. 
In this way, we better train our model, and the model
learns things more effectively.
Table~\ref{tab:architecturecomponentsconfig} summarizes the details of the model architecture.

During testing/prediction,
we feed an unlabeled test data sample to SYMHnet
where the test data sample is the same as the training data sample,
except that the test data sample does not have a label.
The trained SYMHnet model will predict the label based on the input test data sample.
SYMHnet uses
the MC dropout sampling method described in Section \ref{sec:uq}
to produce, for a test data sample, a
predicted SYM-H index value accompanied by results of aleatoric uncertainty and
epistemic uncertainty. 

\begin{table}
\caption{Architecture Details of SYMHnet}
 \label{tab:architecturecomponentsconfig}
 \centering
 \begin{tabular}{ll l }
 \hline
  Component & Parameter & Value    \\ 
 \hline
   Forward LSTM    & Number of LSTM units & 400 \\
   Backward LSTM    & Number of LSTM units & 400 \\
  &Activation function & ReLU\\
  \\
  GNN & Number of nodes & 8 \\ 
  & Number of edges & 56\\
  &Activation function & ReLU\\
  & Number of graph convolutional layers & 2\\
  \\
  Dense layer & Number of neurons & 200 \\
 \hline
 \multicolumn{2}{l}{}
 \end{tabular}
\end{table}

\subsection{Uncertainty Quantification} \label{sec:uq}
Quantification of uncertainty
is essential for the reproducibility and validation of a model
\cite{Volodina2021ImportanceOfUncertaintyQuantification}. 
Uncertainty quantification with deep learning has been used in computer vision \cite{UncertatintyComputervision10.5555/3295222.3295309},  
space weather \cite{GCS2018}, 
and solar physics \cite{HaodiFibri2021}. 
There are two types of uncertainty:
aleatoric and epistemic.
Aleatoric uncertainty
captures the inherent randomness of data,
hence also referred to as data uncertainty. 
Epistemic uncertainty occurs due to
the inexact weight calculations in a neural network
and is also known as model uncertainty. 

In incorporating Bayesian inference into SYMHnet, 
our goal is to find
the posterior distribution over the weights of the network, $W$,
given the observed training data, $X$, and the labels $Y$, that is,
$P(W|X,Y)$.
The posterior distribution is intractable \cite{HaodiFibri2021}, and
one has to approximate the weight distribution
\cite{TransferNNTOProb10.5555/2986766.2986882}.
We use variational inference as suggested by \citeA{Graves_VaritionalNIPS2011_7eb3c8be}
to learn the variational distribution
on the weights of the network, $q(W)$,
by minimizing the Kullback–Leibler (KL) divergence of $q(W)$ and $P(W|X,Y)$.

Training a network with dropout \cite{DropOutJMLR:v15:srivastava14a} is equivalent to a variational approximation
on the network \cite{MoteCarolDropout10.5555/3045390.3045502}.
Furthermore, minimizing
the loss function of cross-entropy (CE)
\cite{Goodfellow_DeepLearningBookDBLP:books/daglib/0040158} can have the same effect as minimizing the
KL divergence term. 
Minimizing CE loss in classification problems is
equivalent to minimizing mean squared error (MSE) loss in regression problems \cite{CrossEntropyMSEUsageExample2020,CrossEntropyMSEComp2005}.
Therefore, we use the MSE loss function and
the root mean squared propagation (RMSProp) optimizer with a learning rate of 0.0002
to train SYMHnet.
Table \ref{tab:modelhyperparameters} summarizes
the hyperparameters and their values used by SYMHnet.
We use $\hat{q}(W)$ to represent the optimized weight
distribution.

During testing/prediction, SYMHnet uses
the MC dropout sampling method 
\cite{MoteCarolDropout10.5555/3045390.3045502} 
to quantify uncertainty.
Specifically, 
we process the test data $K$ times to generate $K$ MC samples
where $K$ is set to 100.
We have experimented with different $K$ values. 
Using a $K$ value of less than 100 does not generate enough samples; the produced uncertainty ranges are too large to be useful. 
Using a $K$ value of larger than 100 increases computation time, while the model performance remains the same. As a consequence, we set $K$ to 100
to process the test data 100 times.
Each time, a set of weights is randomly drawn from $\hat{q}(W)$.
We obtain the mean and variance for the $K$ samples. The mean is the anticipated SYM-H value. 
According to~\citeA{HaodiFibri2021}, we split the variance into aleatoric and epistemic uncertainties.

\begin{table}
\caption{Hyperparameter Values Used by SYMHnet}
 \label{tab:modelhyperparameters}
 \centering
 \begin{tabular}{l l }
 \hline
  Parameter & Value    \\ 
 \hline
  Dropout rate & 0.5\\
  Batch size&1024\\
  Epochs & 50 \\
  Optimizer & RMSProp\\
   Learning rate &   0.0002    \\
 Loss function & MSE \\
 \hline
 \multicolumn{2}{l}{}
 \end{tabular}
\end{table}

\section{Experiments and Results}\label{sec:results}
\subsection{Performance Metrics}\label{sec:evaluationmetrics}
\label{sec:performancemetrics}
To assess the prediction accuracy of SYMHnet and compare it with related machine learning models, 
we adopt
the following metrics: 
root mean square error (RMSE),
forecast skill score (FSS) and
R-squared (R\textsuperscript{2}).
These metrics have been used in the forecasting of geomagnetic indices and are recommended in the literature
\cite{SurvyMLComporeale2019R2,
2022XGBoostSYMHbyIong,
ModelMetricsGuideline2018Liemohn}.
Our experiments were carried out by feeding time series data samples
from the training storms in Table \ref{tab:trainingstorms}
(training set)
to train a model.
We then used the time series data samples
from the validation storms
in Table \ref{tab:validationstorms}
(validation set)
to validate the model and optimize its hyperparameters.
Finally, we used the trained model to predict
the SYM-H index values of
the time series data samples
from the test storms
in Table \ref{tab:teststorms}
(test set).

RMSE measures the difference between prediction and ground truth for each test data sample. It is calculated as follows:\\	\begin{equation}\label{eq:rmse}
		\mbox{RMSE} = \sqrt{\frac{1}{n}\sum_{i=1}^{n}({y_i} - \hat{y_i})^2},
	\end{equation}
	where $n$ is the number of test data samples in a test storm in Table \ref{tab:teststorms}, and
$\hat{y_i}$ ($y_i$, respectively) represents the predicted SYM-H index value (observed SYM-H index value, respectively) at the time point $i$
 in the test storm. 
	The smaller the RMSE, the more accurate the model.
 
 FSS is calculated using the prediction provided by the Burton equation \cite{OBrienBurton2000} as a baseline and is defined as follows
\cite{2022XGBoostSYMHbyIong,Murphy1988MWRv..116.2417MUsingBurton}:\\
\begin{equation}\label{eq:fss}
\mbox{FSS} = 1 - \frac{\frac{1}{n}\sum_{i=1}^{n}({y_i} - \hat{y_i})^2}{\frac{1}{n}\sum_{i=1}^{n}({y_i} - y_i^b)^2}
\end{equation}
where $y_i^b$ represents the prediction provided by the Burton equation at the time point $i$ in the test storm.  
 The FSS value between 0 and 1 indicates that the model is better than the baseline, while the negative FSS value indicates that the model is worse than the baseline \cite{2022XGBoostSYMHbyIong}.
 
R\textsuperscript{2} 
determines the amount of variance of the observed data explained by the predicted data.
It is calculated as follows:\\
	\begin{equation}\label{eq:r2squared}
		\mbox{
    R\textsuperscript{2}
    } = 1- \frac{\sum_{i=1}^{n} ({y_i} - \hat{y_i})^2}
		{\sum_{i=1}^{n} ({y_i} - \bar{y})^2},
	\end{equation}
 where $\bar{y}$ is the mean of the observed SYM-H index values for the test data samples in the test storm.
 The larger the R\textsuperscript{2}, the more accurate the model. 

For each metric, the mean and standard deviation of the metric values for all test storms in the test set (Table \ref{tab:teststorms}) are calculated and recorded.
 
\subsection{Results Based on 1-Minute Resolution Data}
\label{sec:1minuteresult}
In this section, we present experimental results based on the 1-minute resolution data in our database. 
First, we conducted an ablation study to
analyze and assess the components of SYMHnet.
Then we performed 
case studies on a moderately large storm (storm \#36 with SYM-H = $-137$ nT)
and a very large storm (storm \#37 with SYM-H = $-393$ nT)
in the test set
shown in Table \ref{tab:teststorms}
where both storms were previously investigated by 
\citeA{2022XGBoostSYMHbyIong}.
It should be noted that the work of \citeA{2022XGBoostSYMHbyIong} 
was based on 5-minute resolution data.
To our knowledge, no previous method used 1-minute resolution data
to predict the SYM-H index.
\vspace{-0.18cm}
\subsubsection{Ablation Study with 1-Minute Resolution Data}\label{sec:ablation1min}

We considered three variants of SYMHnet:
SYMHnet-B, SYMHnet-G and SYMHnet-BG. 
SYMHnet-B represents the subnetwork of SYMHnet 
with the BiLSTM component removed. 
SYMHnet-G represents the subnetwork of SYMHnet 
with the GNN component removed.
SYMHnet-BG represents the subnetwork of SYMHnet
with both the BiLSTM and GNN components removed.
Thus, SYMHnet-BG simply contains the dense layers in SYMHnet, 
which amounts to a simple multilayer perceptron network.
When conducting the ablation study,
we turned off the uncertainty quantification mechanism.

Table~\ref{tab:symh_all_metrics_1minutes_ablation} presents
the average values for RMSE, FSS, and R\textsuperscript{2} 
(with standard deviations enclosed in parentheses)
obtained by the four models: 
SYMHnet,
SYMHnet-B, 
SYMHnet-G and SYMHnet-BG,
based on the 1-minute resolution data in our database.
The best metric values are highlighted in boldface.
It can be seen from 
Table~\ref{tab:symh_all_metrics_1minutes_ablation}
that SYMHnet outperforms its three variants.  
SYMHnet-B is the second best among the four models, implying that a GNN is effective in solving time series regression problems~\cite{GNNForTimeSeries2022Stefan}. 
SYHMnet-G, which contains a BiLSTM network but no GNN, does not perform well.
This finding is consistent with those in \citeA{2021ColladoSMYH_ASYH_CNN_LSTMForecasting}, who showed that
LSTM performed worse than a combination of LSTM and CNN in
SYM-H forecasting.
Finally, SYMHnet-BG is the worst among the four models.
This happens because SYMHnet-BG loses the advantages offered by GNN and BiLSTM networks.

\begin{table}
    \centering
       \caption{Results of the Ablation Study Based on 1-Minute Resolution Data}
    \begin{tabular}{c c r r r r}
     \hline
Metric & Hour-ahead  & SYMHnet &  SYMHnet-B & SYMHnet-G & SYMHnet-BG  \\
 \hline
 RMSE &
1 & \textbf{3.002} (2.169) & 3.210 (2.319) & 4.194 (3.030) & 5.348 (2.957) \\
& 2 & \textbf{3.171} (2.201) & 3.432 (2.382) & 4.369 (3.033) & 5.623 (3.066) \\
\hline
FSS &1 & \textbf{0.668} (0.131) & 0.563 (0.003) & 0.007 (0.012) & $-0.644$ (0.015) \\
    &2 & \textbf{0.760} (0.089) & 0.387 (0.031) & $-0.367$ (0.016) & $-0.731$ (0.031) \\
    \hline
R$^2$ 
 & 1 & \textbf{0.993} (0.003) & 0.913 (0.001) & 0.789 (0.001) & 0.602 (0.001) \\
 & 2 & \textbf{0.993} (0.003) & 0.908 (0.002) & 0.776 (0.002) & 0.594 (0.002) \\
\hline    
    \end{tabular}
    \label{tab:symh_all_metrics_1minutes_ablation}
\end{table}

\subsubsection{Case Studies with 1-Minute Resolution Data}\label{sec:casestudy1min}

\begin{figure}[ht!]
    \centering
    \includegraphics[width=1\textwidth]{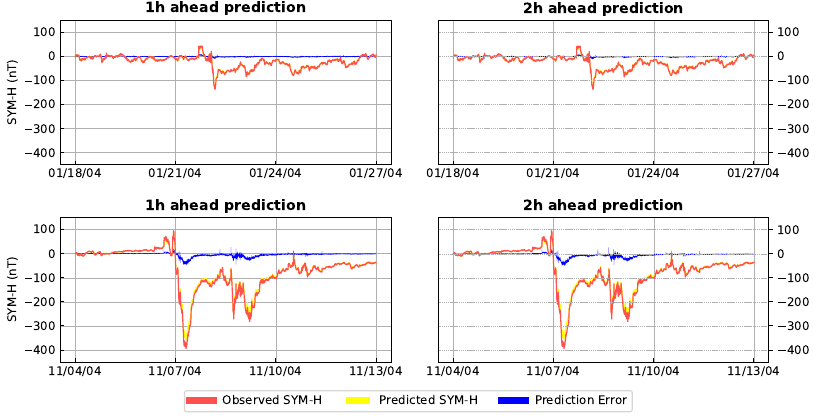}
    \caption{Predictions for storm \#36 (top) and storm \#37 (bottom) made by the SYMHnet model
    based on 1-minute resolution data.
    The red line represents the observed SYM-H values, 
    the yellow dashed line represents the model's predictions, 
    and the blue line represents the prediction error.
    Both quiet time and storm time are shown in the figure.
    }
    \label{fig:pe_37_36_1min}
\end{figure}

Here we conducted case studies by using SYMHnet to predict
the SYM-H index values in storms \#36 
and 
\#37 given in Table \ref{tab:teststorms} based on the 
1-minute resolution data in our database.
Additional case studies on other storms can be found in Appendix A.
The period of storm \#36 started on 18 January 2004 and ended on 27 January 2004, 
with a minimum SYM-H value of $-137$ nT
and a maximum SYM-H value of 41 nT
during the period.
The period of storm \#37 started on 4 November 2004 and ended on 14 November 2004, 
with a minimum SYM-H value of $-393$ nT
and a maximum SYM-H value of 92 nT
during the period. 
Figure~\ref{fig:pe_37_36_1min} 
shows the predictions and measured error of the SYMHnet model in storm \#36 and storm \#37 respectively. 
In the figure, each point on a yellow dashed line represents the prediction made at the corresponding time $x$ on the X-axis. 
For 1-hour ahead (2-hour ahead, respectively) predictions,
the point/prediction at time $x$ is produced based on the solar wind/IMF parameters at time $x$ – 1 hour ($x$ – 2 hours, respectively). 
There is a lag of 1 hour (for 1-hour ahead predictions) or 2 hours (for 2-hour ahead predictions) as in previous studies \cite{2021ColladoSMYH_ASYH_CNN_LSTMForecasting,2022XGBoostSYMHbyIong}.
It can be seen from Figure~\ref{fig:pe_37_36_1min} that
the SYMHnet model works well at both quiet time and storm time.
The measured error ranges between $-15$ nT and 23 nT for storm \#36
and between $-50$ nT and 34 nT for storm \#37.
The more intense the storm, the larger the measured error.

\begin{figure}[ht!]
    \centering
    \includegraphics[width=1\textwidth]{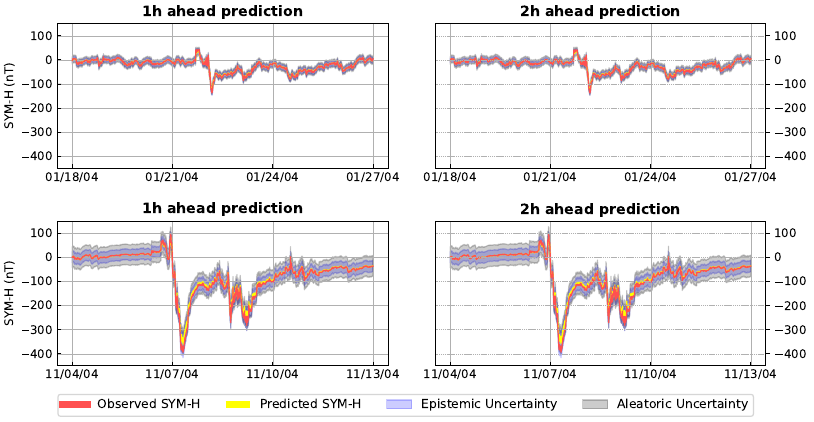}
    \caption{Uncertainty quantification results 
    produced by the SYMHnet model 
    in storm \#36 (top) and storm \#37 (bottom) based on 1-minute resolution data. 
    The red line represents the observed SYM-H values,
    the yellow dashed line represents the model's predictions,
     the light-blue region shows epistemic uncertainty (model uncertainty), 
    and the light-gray region shows aleatoric uncertainty (data uncertainty).
    Both quiet time and storm time are shown in the figure.}
    \label{fig:uncertaintysymh_storm_36_3_6_1min}
\end{figure}

Figure~\ref{fig:uncertaintysymh_storm_36_3_6_1min} 
presents uncertainty quantification results produced by SYMHnet 
in storm \#36 and storm \#37, respectively,
based on the 1-minute resolution data in our database.
In the figure, the red line represents the observed values of the SYM-H index,
and the yellow dashed line represents the predicted values of the SYM-H index.
The light-blue region shows the epistemic uncertainty (model uncertainty) and
the light-gray region shows the aleatoric uncertainty (data uncertainty)
of the predicted outcome.
It can be seen in Figure~\ref{fig:uncertaintysymh_storm_36_3_6_1min} that the 
yellow dashed line (predicted values) 
is reasonably close to the red line (observed values), again demonstrating the good performance of SYMHnet. 
The light-blue region is tinier than the 
light-gray region, 
indicating that the model uncertainty is lower than the data uncertainty. 
This is due to the fact that the uncertainty in the predicted outcome
is primarily caused by the noise in the input test data, not by the SYMHnet model.

\subsection{Results Based on 5-Minute Resolution Data}
\label{sec:5minuteresult}

SYMHnet can be easily modified to process 5-minute resolution data.
As described in Section \ref{sec:architecture},
the input data sample at the time point $t$ is composed of the parameter graph $G_{t}$
and a sequence of $m$ records.
The difference is that the cadence of the $m$ records here is 5-minute 
rather than 1-minute.
Furthermore, the labels of the parameter graphs $G_{t}$, $G_{t+5}$,
$G_{t+10}$ are the SYM-H index values
at the time points $t$ + $w$ hour,
($t$ + 5 minutes) + $w$ hour,
($t$ + 10 minutes) + $w$ hour, respectively,
for $w$-hour 
ahead predictions where $w$ is 1 or 2.

In the following, we present experimental results based on the 5-minute resolution data
in our database.
As in Section \ref{sec:1minuteresult}, we
conducted an ablation study,
this time using the 5-minute resolution data.
We then performed case studies on 
storms \#36 and \#37.
Finally, we compared SYMHnet with related machine learning methods,
all of which utilized the 5-minute resolution data in our database.
Since the related methods cannot quantify uncertainty,
we turned off the uncertainty quantification mechanism in SYMHnet
while conducting the comparative study.

\subsubsection{Ablation Study with 5-Minute Resolution Data}\label{sec:ablation5min}

Table~\ref{tab:symh_all_metrics_5minutes_ablation} presents 
the average values for 
RMSE, FSS and R\textsuperscript{2} 
(with standard deviations enclosed in parentheses) 
obtained by the four models:
SYMHnet,
SYMHnet-B, SYMHnet-G and SYMHnet-BG,
based on the 5-minute resolution data in our database.
The best metric values are highlighted in boldface.
It can be seen from 
Table \ref{tab:symh_all_metrics_5minutes_ablation} that SYMHnet 
is again the best among the four models for the 5-minute resolution data,
a finding consistent with that in Table \ref{tab:symh_all_metrics_1minutes_ablation}
for the 1-minute resolution data.
\begin{table}
    \centering
       \caption{
       Results of the Ablation Study Based on 5-Minute Resolution Data}
    \begin{tabular}{c c r r r r}
     \hline
Metric & Hour-ahead  & SYMHnet &  SYMHnet-B & SYMHnet-G & SYMHnet-BG  \\
 \hline
 RMSE
& 1 &  \textbf{5.914} (2.169) & 6.324 (2.319) & 8.262 (2.834) & 10.537 (2.958) \\
& 2 & \textbf{6.481} (2.201) & 8.646 (2.636) & 13.021 (3.315) & 14.165 (3.194) \\
\hline
FSS 
 & 1 & \textbf{0.484} (0.195) & 0.407 (0.087) & 0.005 (0.012) & $-0.465$ (0.023) \\
 & 2 & \textbf{0.593} (0.096) & 0.302 (0.048) & $-0.286$ (0.035) & $-0.570$ (0.042) \\
    \hline
R$^2$ 
  & 1 & \textbf{0.993} (0.003) & 0.912 (0.003) & 0.789 (0.004) & 0.601 (0.005)  \\
  & 2 & \textbf{0.989} (0.003) & 0.905 (0.003) & 0.773 (0.004) & 0.592 (0.005)  \\
\hline    
    \end{tabular}
    \label{tab:symh_all_metrics_5minutes_ablation}
\end{table}

\subsubsection{Case Studies with 5-Minute Resolution Data}

\begin{figure}
    \centering
    \includegraphics[width=1\textwidth]{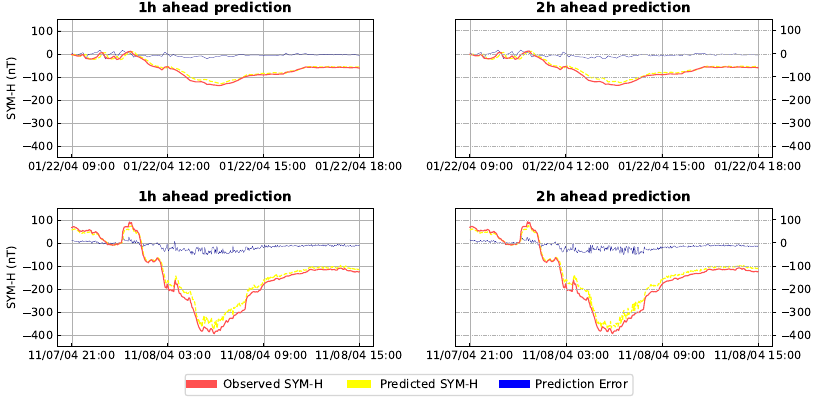}
    \caption{Predictions for storm \#36 (top) and storm \#37 (bottom)
    made by the SYMHnet model based on 5-minute resolution data.
    The red line represents the observed SYM-H values,
    the yellow dashed line represents the model's predictions, and
    the blue line represents the prediction error.
    Only the peak storm time is shown in the figure.
    }
    \label{fig:pe_36_37_5min}
\end{figure}
\begin{figure}[ht!]
    \centering  
    \includegraphics[width=1\textwidth]{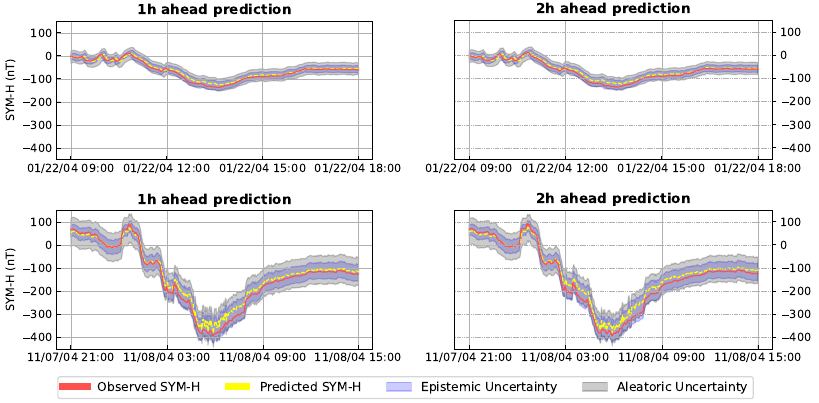}
    \caption{Uncertainty quantification results produced by the SYMHnet model in
    storm \#36 (top) and storm \#37 (bottom) based on 5-minute resolution data.
    The red line represents the observed SYM-H values, 
    the yellow dashed line represents the model’s predictions, 
    the light-blue region 
    shows epistemic uncertainty (model uncertainty), and the 
    light-gray region shows aleatoric uncertainty (data uncertainty).  
     Only the peak storm time is shown in the figure.
    }
    \label{fig:uncertaintysymh_storm_36_37_hrs_5min_lv}
\end{figure}

Figure \ref{fig:pe_36_37_5min} shows the predictions and measured error of 
SYMHnet in storms 
\#36 and \#37, respectively, and
Figure~\ref{fig:uncertaintysymh_storm_36_37_hrs_5min_lv} 
presents the uncertainty quantification results
produced by SYMHnet in these storms respectively,
based on the 5-minute resolution data in our database.
Unlike Figures \ref{fig:pe_37_36_1min} and \ref{fig:uncertaintysymh_storm_36_3_6_1min},
in which both quiet time and storm time are shown,
Figures \ref{fig:pe_36_37_5min} and \ref{fig:uncertaintysymh_storm_36_37_hrs_5min_lv}
focus on the peak storm time.
In Figure \ref{fig:pe_36_37_5min}, 
the measured error ranges 
between $-24$ nT and 25 nT for storm \#36 and
between $-52$ nT and 36 nT for storm \#37.
These results indicate that SYMHnet 
can properly forecast the SYM-H index 
even in the most intense storm period.

In Figure~\ref{fig:uncertaintysymh_storm_36_37_hrs_5min_lv},
the red line represents the
observed values of the SYM-H and the 
yellow dashed line represents the predicted
values of the SYM-H. 
The light-blue area shows the epistemic uncertainty (model uncertainty) and
the light-gray area shows the aleatoric uncertainty (data uncertainty)
of the predicted outcome.
It can be seen from 
Figure~\ref{fig:uncertaintysymh_storm_36_37_hrs_5min_lv} 
that the red line representing the observed 
SYM-H values is within the uncertainty interval, 
indicating SYMHnet's predicted values together with
the uncertainty values well cover the observed values.
The overall findings here are similar to those from the 1-minute resolution data 
shown in Figure~\ref{fig:uncertaintysymh_storm_36_3_6_1min}.

\subsubsection{Comparative Study with 5-Minute Resolution Data}
\label{sec:modelscomparison}
Several researchers performed SYM-H forecasting using machine learning and the 5-minute resolution data.
\citeA{2021ColladoSMYH_ASYH_CNN_LSTMForecasting} combined 
long short-term memory (LSTM) and a convolutional neural network (CNN),
referred to as the LCNN method, to forecast the SYM-H index (1 and 2 hours in advance).
\citeA{2022XGBoostSYMHbyIong} utilized 
gradient boosting machines, referred to as the GBM method, to
forecast the SYM-H index (also 1 and 2 hours in advance). \citeA{SYMHForecasting2021ComparisonLSTMCNN} 
compared LSTM and CNN for the prediction of the SYM-H index 
(only 1 hour in advance).  
Although the methods including ours use slightly different data samples, these methods are all developed to predict the SYM-H index values in the same set of storms. 
The purpose of this comparative study is to compare the prediction results/accuracies of, rather than specific models/data samples in, these methods. This comparison methodology has commonly been used in  
SYM-H forecasting  \cite{2021ColladoSMYH_ASYH_CNN_LSTMForecasting,2022XGBoostSYMHbyIong,SYMHForecasting2021ComparisonLSTMCNN}.
Since the related methods cannot predict uncertainties, we 
turned off the uncertainty quantification component in SYMHnet
 while carrying out the comparative study.
 The Burton equation \cite{OBrienBurton2000}, used as the baseline,
 is also included.
 The performance metric values of each method for each test storm
 in the test set (Table \ref{tab:teststorms}) are calculated.
 The best metric values are highlighted in boldface.

Tables~\ref{tab:symhrmse_1hr_5min} and \ref{tab:symhrmse_2hr_5min} 
compare the RMSE results of these methods
for 1-hour and 2-hour ahead SYM-H predictions, respectively,
based on the RMSE values available in the related studies 
\cite{2021ColladoSMYH_ASYH_CNN_LSTMForecasting,2022XGBoostSYMHbyIong,OBrienBurton2000,SYMHForecasting2021ComparisonLSTMCNN}.
Tables \ref{tab:symhfss_1hr_5min} and \ref{tab:symhfss_2hr_5min}
compare the FSS results of these methods
for 1-hour and 2-hour ahead SYM-H predictions, respectively,
based on the FSS values available in the related studies
\cite{2021ColladoSMYH_ASYH_CNN_LSTMForecasting,2022XGBoostSYMHbyIong,SYMHForecasting2021ComparisonLSTMCNN}.
Table~\ref{tab:symhr2_1hr_5min} compares the R$^2$ results
of these methods
for 1-hour ahead and 2-hour ahead SYM-H predictions, respectively,
on the same test storms.
\citeA{2022XGBoostSYMHbyIong} did not provide R$^2$ results, 
and hence the GBM method was excluded from Table~\ref{tab:symhr2_1hr_5min}.
These tables show that SYMHnet performs better than the related methods
for all except two test storms (\#28 and/or \#40),
demonstrating the good performance and feasibility of our tool for SYM-H forecasting.

\begin{table}
    \centering
       \caption{RMSEs for 1-hr Ahead Prediction from 
       the Comparative Study Including 
       SYMHnet, 
       LCNN \cite{2021ColladoSMYH_ASYH_CNN_LSTMForecasting}, 
       GBM \cite{2022XGBoostSYMHbyIong}, LSTM and CNN \cite{SYMHForecasting2021ComparisonLSTMCNN}, 
       and Burton Equation \cite{OBrienBurton2000}}
    \begin{tabular}{c r r r r r r}
    \hline
    & \multicolumn{5}{c}{1-h ahead prediction (RMSE)} \\
     \hline
Storm \#  & SYMHnet &  LCNN & GBM & LSTM & CNN & Burton \\
 \hline
26  & \textbf{3.977} &  6.630  &  5.863  &  6.700  &  7.200  &  6.839  \\
27  & \textbf{7.682} &  8.913  &  7.729  &  8.900  &  10.500  &  7.955  \\
28  & 4.599 &  5.858  &  \textbf{4.281}  &  5.400  &  5.600  &  5.967  \\
29  & \textbf{5.058} &  6.683  &  5.833  &  7.200  &  7.700  &  6.511  \\
30  & \textbf{2.213} &  5.200  &  4.927  &  5.600  &  6.500  &  4.614  \\
31  & \textbf{7.923} &  8.584  &  8.277  &  10.700  &  9.600  &  8.838  \\
32  & \textbf{3.969} &  7.259  &  6.841  &  8.300  &  8.200  &  9.487  \\
33  & \textbf{11.366} &  13.340  &  14.492  &  16.300  &  19.100  &  16.630  \\
34  & \textbf{5.259} &  10.034  &  10.190  &  11.300  &  12.400  &  10.888  \\
35  & \textbf{5.406} &  7.693  &  7.154  &  8.500  &  8.800  &  7.918  \\
36  & \textbf{5.618} &  9.525  &  8.512  &  8.700  &  10.500  &  9.082  \\
37  & \textbf{10.320} &  15.184  &  14.548  &  17.500  &  17.300  &  15.713  \\
38  & \textbf{3.368} &  4.080  &  3.886  &  4.200  &  4.600  &  4.572  \\
39  & \textbf{5.670} &  6.431  &  5.901  &  5.700  &  6.800  &  6.663  \\
40  & 5.752 &  \textbf{4.673}  &  4.976  &  5.500  &  5.900  &  5.371  \\
41  & \textbf{5.871} &  7.882  &  7.558  &  9.000  &  9.400  &  8.358  \\
42  & \textbf{3.900} &  5.669  &  5.030  &  5.900  &  6.300  &  5.549  \\
\hline
    \end{tabular}
    \label{tab:symhrmse_1hr_5min}
\end{table}

\begin{table}
    \centering
    \caption{RMSEs for 2-hr Ahead Prediction 
    from the Comparative Study Including SYMHnet,
    LCNN \cite{2021ColladoSMYH_ASYH_CNN_LSTMForecasting}, 
    GBM \cite{2022XGBoostSYMHbyIong},  
    and Burton Equation~\cite{OBrienBurton2000}}
    \begin{tabular}{c r r r r}
    \hline
    & \multicolumn{3}{c}{2-h ahead prediction (RMSE)} \\
     \hline
Storm \#  & SYMHnet &  LCNN & GBM & Burton \\
 \hline
26  & \textbf{4.330} &  8.989  &  8.285  & 10.690\\
27  & \textbf{8.577} &   13.418  &  11.585  &  12.465 \\
28  & \textbf{4.977} &  5.877  &  5.650  &  8.858 \\
29  & \textbf{5.515} &  9.314  &  8.826  &  9.776\\
30  & \textbf{2.636} &  7.288  &  7.280  &  6.266 \\
31  & \textbf{9.737} &  12.436  &  12.613  &  13.604 \\
32  & \textbf{4.451} &  8.937  &  9.927  &  13.766 \\
33  & \textbf{13.745} &  18.481  &  24.519  &  25.729 \\
34  & \textbf{5.611} &  13.941  &  13.736  &  14.695 \\
35  & \textbf{5.830} &  9.932  &  9.504  &  10.586 \\
36  & \textbf{5.970} &  12.058  &  12.068  &  13.117 \\
37  & \textbf{10.923} &  21.084  &  22.327  &  24.446 \\
38  & \textbf{3.765} &  5.213  &  5.153  &  6.546 \\
39  & \textbf{6.252} &  6.798  &  7.391  &  10.159 \\
40  & 6.336 &   \textbf{5.281}  &  5.633  &  6.032 \\
41  & \textbf{6.857} &  11.707  &  12.121  &  12.622 \\
42  & \textbf{4.674} &  8.273  &  7.976  &  8.877\\
    \hline
\end{tabular}
    \label{tab:symhrmse_2hr_5min}
\end{table}

\begin{table}
    \centering
    \caption{
    FSSs for 1-hr Ahead Prediction from the Comparative Study Including SYMHnet, 
    LCNN \cite{2021ColladoSMYH_ASYH_CNN_LSTMForecasting}, GBM \cite{2022XGBoostSYMHbyIong}, 
    LSTM and CNN \cite{SYMHForecasting2021ComparisonLSTMCNN}}
    \begin{tabular}{c r r r r r}
        \hline
    & \multicolumn{5}{c}{1-h ahead prediction (FSS)} \\
     \hline
Storm \# & SYMHnet & LCNN & GBM & LSTM & CNN  \\
\hline
26 & \textbf{0.418} & 0.031 & 0.143 & 0.020 & $-0.053$ \\
27 & \textbf{0.034} & $-0.120$ & 0.028 & $-0.119$ & $-0.320$ \\
28 & 0.229 & $-0.028$ & \textbf{0.249} & 0.095 & 0.062 \\
29 & \textbf{0.223} & $-0.026$ & 0.104 & $-0.106$ & $-0.183$ \\
30 & \textbf{0.520} & $-0.127$ & $-0.068$ & $-0.214$ & $-0.409$ \\
31 & \textbf{0.104} & 0.029 & 0.063 & $-0.211$ & $-0.086$ \\
32 & \textbf{0.582} & 0.235 & 0.279 & 0.125 & 0.136 \\
33 & \textbf{0.317} & 0.198 & 0.129 & 0.020 & $-0.149$ \\
34 & \textbf{0.517} & 0.078 & 0.064 & $-0.038$ & $-0.139$ \\
35 & \textbf{0.317} & 0.028 & 0.096 & $-0.074$ & $-0.111$ \\
36 & \textbf{0.381} & $-0.049$ & 0.063 & 0.042 & $-0.156$ \\
37 & \textbf{0.343} & 0.034 & 0.074 & $-0.114$ & $-0.101$ \\
38 & \textbf{0.263} & 0.108 & 0.150 & 0.081 & $-0.006$ \\
39 & \textbf{0.149} & 0.035 & 0.114 & 0.160 & $-0.021$ \\
40 & $-0.071$ & \textbf{0.130} & 0.074 & $-0.024$ & $-0.098$ \\
41 & \textbf{0.298} & 0.057 & 0.096 & $-0.077$ & $-0.125$ \\
42 & \textbf{0.297} & $-0.022$ & 0.094 & $-0.063$ & $-0.135$ \\
         \hline
    \end{tabular}
    \label{tab:symhfss_1hr_5min}
\end{table}

\begin{table}
    \centering
    \caption{
    FSSs for 2-hr Ahead Prediction from the Comparative Study Including 
    SYMHnet, LCNN \cite{2021ColladoSMYH_ASYH_CNN_LSTMForecasting}, and GBM \cite{2022XGBoostSYMHbyIong}}
    \begin{tabular}{c r r r}
    \hline
    & \multicolumn{3}{c}{2-h ahead prediction (FSS)} \\
     \hline    
Storm \# & SYMHnet  & LCNN & GBM  \\
\hline
26 & \textbf{0.595} & 0.159 & 0.225 \\
27 & \textbf{0.312} & $-0.076$ & 0.071 \\
28 & \textbf{0.438} & 0.337 & 0.362 \\
29 & \textbf{0.436} & 0.047 & 0.097 \\
30 & \textbf{0.579} & $-0.163$ & $-0.162$ \\
31 & \textbf{0.284} & 0.086 & 0.073 \\
32 & \textbf{0.677} & 0.351 & 0.279 \\
33 & \textbf{0.466} & 0.282 & 0.047 \\
34 & \textbf{0.618} & 0.051 & 0.065 \\
35 & \textbf{0.449} & 0.062 & 0.102 \\
36 & \textbf{0.545} & 0.081 & 0.080 \\
37 & \textbf{0.553} & 0.138 & 0.087 \\
38 & \textbf{0.425} & 0.204 & 0.213 \\
39 & \textbf{0.385} & 0.331 & 0.272 \\
40 & $-0.050$ & \textbf{0.125} & 0.066 \\
41 & \textbf{0.457} & 0.072 & 0.040 \\
42 & \textbf{0.473} & 0.068 & 0.101 \\
         \hline
    \end{tabular}
    \label{tab:symhfss_2hr_5min}
\end{table}

\begin{table}
    \centering
    \caption{R$^2$s for 1- and 2-hr Ahead Predictions 
     from the Comparative Study Including SYMHnet,
    LCNN \cite{2021ColladoSMYH_ASYH_CNN_LSTMForecasting}, 
    LSTM and CNN \cite{SYMHForecasting2021ComparisonLSTMCNN}}
    \begin{tabular}{c r r r r| r r}
    \hline
    & \multicolumn{4}{c}{1-h ahead prediction (R$^2$)} & 
    \multicolumn{2}{|c}{2-h ahead prediction  (R$^2$)}\\
     \hline
Storm \#  &  SYMHnet  &  LCNN  &  LSTM  &  CNN  &      SYMHnet  &  LCNN\\
\hline
26  & \textbf{0.956} & 0.870 &  0.890  &  0.870  &       \textbf{0.948} &   0.766\\
27  & \textbf{0.952} & 0.939 &  0.940  &  0.920  &       \textbf{0.940} &   0.862\\
28  & \textbf{0.957} & 0.936 &  0.950  &  0.950  &       \textbf{0.949} &   0.936\\
29  & \textbf{0.955} & 0.922 &  0.930  &  0.920  &       \textbf{0.946} &   0.848\\
30  & \textbf{0.991} & 0.946 &  0.950  &  0.930  &       \textbf{0.987} &   0.894\\
31  & \textbf{0.976} & 0.971 &  0.960  &  0.970  &       \textbf{0.963} &   0.939\\
32  & \textbf{0.986} & 0.953 &  0.950  &  0.950  &       \textbf{0.982} &   0.929\\
33  & \textbf{0.979} & 0.965 &  0.960  &  0.950  &       \textbf{0.969} &   0.932\\
34  & \textbf{0.945} & 0.798 &  0.750  &  0.700  &       \textbf{0.938} &   0.612\\
35  & \textbf{0.954} & 0.907 &  0.900  &  0.890  &       \textbf{0.947} &   0.845\\
36  & \textbf{0.949} & 0.864 &  0.890  &  0.840  &       \textbf{0.943} &   0.782\\
37  & \textbf{0.993} & 0.966 &  0.960  &  0.960  &       \textbf{0.992} &   0.934\\
38  & \textbf{0.961} & 0.939 &  0.940  &  0.930  &       \textbf{0.951} &   0.900\\
39  & \textbf{0.964} & 0.932 &  0.960  &  0.940  &       \textbf{0.943} &   0.924\\
40  & 0.948 & \textbf{0.966} &  0.950  &  0.950  &       0.937 &   \textbf{0.957}\\
41  & \textbf{0.984} &  0.969  &  0.960  &  0.960  &       \textbf{0.978} &   0.931\\
42  & \textbf{0.985} &  0.968  &  0.970  &  0.960  &       \textbf{0.978} &   0.932\\
\hline
    \end{tabular}
    \label{tab:symhr2_1hr_5min}
\end{table}

\section{Discussion and Conclusion}
\label{sec:conclusion}
Geomagnetic activities have a significant impact on Earth, 
which can cause damages to spacecraft, electrical power grids, and navigation systems.
Geospace scientists use geomagnetic indices to measure and quantify the geomagnetic activities. 
The SYM-H index provides information about the response and behavior of the Earth's magnetosphere 
during geomagnetic storms. 
Therefore, a lot of effort has been put into SYM-H forecasting.
Previous work mainly focused on 5-minute resolution data and
skipped 1-minute resolution data.
The higher temporal resolution of the 1-minute resolution data poses a more difficult challenge to forecast due to its highly oscillating character.
This oscillating behavior could make the data more noisy to a machine learning model. 
As a consequence, the model requires more iterations during training with a larger number of neurons in order to learn more features and patterns hidden in the data.

In our study, the SYMHnet model architectures for processing the 1-minute resolution data 
and 5-minute resolution data are the same,
as shown in Figure~\ref{fig:architecture}.
The configuration details and hyperparameter values of SYMHnet for processing the 5-minute resolution data 
are shown in
Tables \ref{tab:architecturecomponentsconfig} and \ref{tab:modelhyperparameters}.
When processing the 1-minute resolution data,
the model is configured with a larger number of neurons
in the dense layers, a higher percentage in the dropout layers,
and a larger number of epochs during the training phase.
This configuration is designed to combat the highly oscillating behavior of the 1-minute resolution data.

Results from our experiments demonstrated the 
good performance of SYMHnet
at both quiet time and storm time.
These results were obtained from a 
database of 42 storms that occurred 
between 1998 and 2018
during the past two solar cycles (\#23 and \#24). 
As done in previous studies
\cite{2021ColladoSMYH_ASYH_CNN_LSTMForecasting,
2022XGBoostSYMHbyIong,
SYMHForecasting2021ComparisonLSTMCNN},
20 storms, listed in Table \ref{tab:trainingstorms}, were used for training,
5 storms, listed in Table \ref{tab:validationstorms},
were used for validation,
and 17 storms, listed in Table \ref{tab:teststorms},
were used for testing.
Based on the tables, the 42 storms were distributed to 14 distinct years.

To avoid bias in drawing a conclusion from the above experiments, 
we conducted an additional experiment 
using 14-fold cross validation
where the data was divided into 14 partitions or folds. 
Each fold corresponds to one year in which at least one storm occurred.
The sequential order of the data in each fold was maintained.
In each run, one fold was used for testing and the other 13 folds together 
were used for training.
Thus, the training set and test set are disjoint, 
and the trained model can predict unseen SYM-H values in the test set.
There were 14 folds and consequently 14 runs 
where the average performance metric values over the 14 runs were calculated.
The results of the 14-fold cross validation were consistent with those reported in the paper.
These results indicate that the SYMHnet tool can be used to predict future SYM-H index values without knowing whether a storm is going to start.
When the predicted SYM-H value is less than a threshold (e.g., $-30$ nT),
the tool detects the occurrence of a storm.
Thus, we conclude that the proposed SYMHnet is a
viable machine learning method
for short-term, 1 or 2-hour ahead forecasts of the SYM-H index for
both 1- and 5-minute resolution data.

\section*{Data Availability Statement}
\label{sec:dataavailability}
\begin{itemize}
    \item 
 The solar wind, IMF and derived parameters along with the SYM-H index data 
used in our study are publicly available from NASA's Space Physics Data Facility at \url{http://omniweb.gsfc.nasa.gov/ow.html}.
 \item {Details of SYMHnet can be found at
\url{https://doi.org/10.5281/zenodo.10602518}.
}

\end{itemize}

\acknowledgments
We appreciate the editor and anonymous referees for constructive comments and suggestions. 
We acknowledge the use of NASA/GSFC's Space Physics Data Facility's OMNIWeb and CDAWeb services, and OMNI data.
This work was supported in part by U.S. NSF grants AGS-1927578, AGS-1954737, AGS-2149748, AGS-2228996,
AGS-2300341 and OAC-2320147.
Huseyin Cavus was supported by the Fulbright Visiting Scholar Program of the Turkish Fulbright Commission.
\clearpage
\appendix
\section{Additional Case Studies with 1-Minute Resolution Data}
\begin{figure}
    \centering
    \includegraphics[width=1\textwidth]{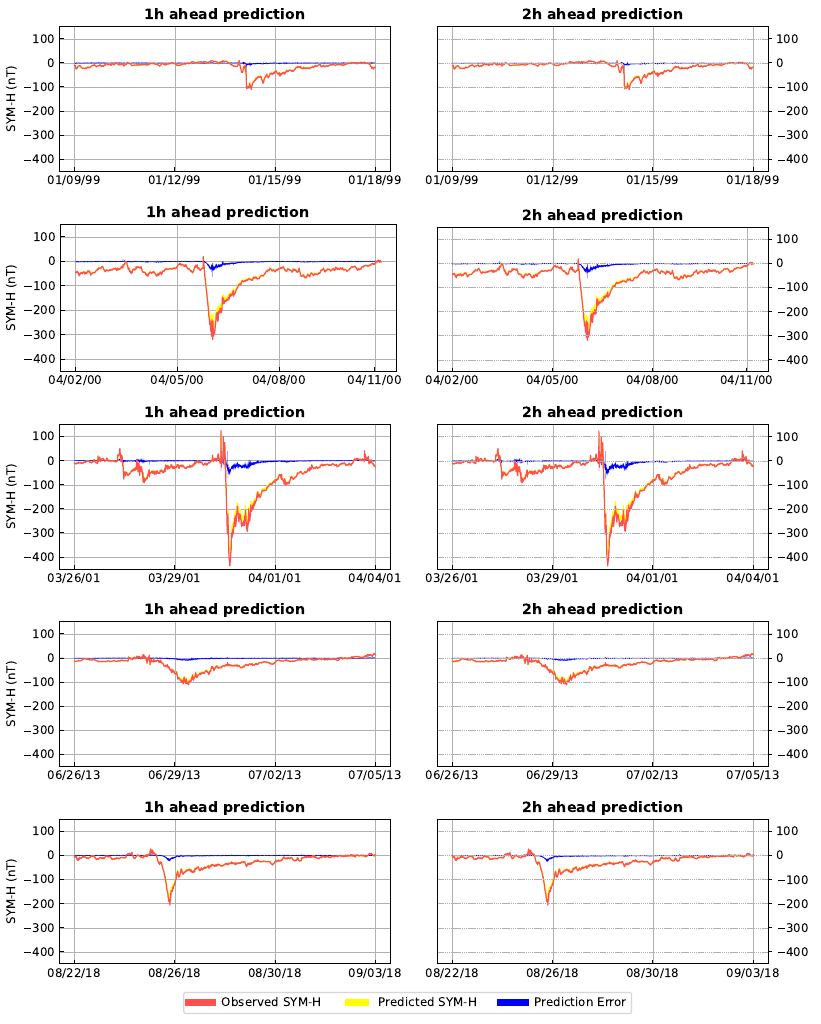}
    \caption{
    Predictions for storms (from top to bottom) \#28 in January 1999, \#31 in April 2000, \#33 in March 2001, \#40 in June 2013, and \#42 in August 2018,
    made by the SYMHnet model based on 1-minute resolution data. 
    The red line represents the observed SYM-H values, 
    the yellow dashed line represents the model's predictions, 
    and the blue line represents the prediction error.
    Both quiet time and storm time are shown in the figure.}
    \label{fig:storms_28_42_1minute_uq_pe}
\end{figure}

\begin{figure}[ht!]
    \centering
    \includegraphics[width=1\textwidth]{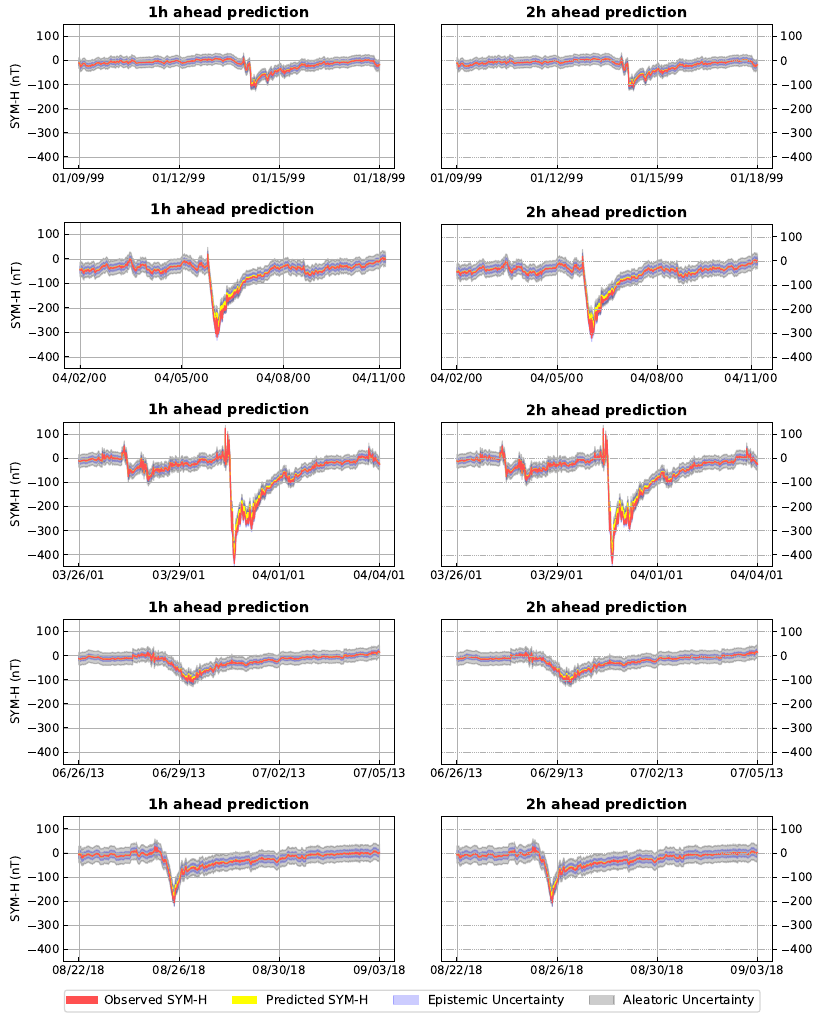}
    \caption{Uncertainty quantification results produced by the SYMHnet model
    in storms (from top to bottom) 
    \#28 in January 1999, 
    \#31 in April 2000, 
    \#33 in March 2001, 
    \#40 in June 2013, and 
    \#42 in August 2018, 
    based on 1-minute resolution data. 
    The red line represents the observed SYM-H values, 
    the yellow dashed line represents the model’s predictions, 
    the light-blue region shows epistemic uncertainty (model uncertainty), 
    and the light-gray region shows aleatoric uncertainty (data uncertainty). 
Both quiet time and storm time are shown in the figure.}
    \label{fig:storms_28_42_1minute_uq}
\end{figure}

Figure \ref{fig:storms_28_42_1minute_uq_pe} shows the predictions and measured error
of SYMHnet in 
storms \#28, \#31, \#33, \#40, and \#42, respectively,
and Figure \ref{fig:storms_28_42_1minute_uq} presents the uncertainty 
quantification results produced by SYMHnet in these storms,
respectively,
based on the 1-minute resolution data in our database. 
The period of storm \#28 started on 9 January 1999 and ended on 18 January 1999, 
with a minimum SYM-H value of $-111$ nT and a maximum SYM-H value of 9 nT. 
The period of storm \#31 started on 2 April 2000 and ended on 12 April 2000, 
with a minimum SYM-H value of $-315$ nT and a maximum SYM-H value of 16 nT. 
The period of storm \#33 stared on 26 March 2001 and ended on 4 April 2001, 
with a minimum SYM-H value of $-434$ nT and a maximum SYM-H value of 109 nT. 
The period of storm \#40 started on 26 June 2013 and ended on 4 July 2013, 
with a minimum SYM-H value of $-110$ nT and a maximum SYM-H value of 19 nT. 
The period of storm \#42 started on 22 August 2018 and ended on 3 September 2018, 
with a minimum SYM-H value of $-205$ nT and a maximum SYM-H value of 26 nT.
In Figure \ref{fig:storms_28_42_1minute_uq_pe}, 
the measured error ranges 
between $-46$ nT and 7 nT for storm \#28,
between $-58$ nT and 2 nT for storm \#31,
between $-69$ nT and 32 nT for storm \#33,
between $-12$ nT and 4 nT for storm \#40, and 
between $-26$ nT and 7 nT for storm \#42.
Generally, the more intense the storm, the larger the measured error.
In Figure \ref{fig:storms_28_42_1minute_uq}, we see that
SYMHnet’s predicted values together with the uncertainty values well cover the observed values, a finding consistent with that in 
Figure \ref{fig:uncertaintysymh_storm_36_3_6_1min}. 

\clearpage

\end{document}